\pdfoutput=1
\documentclass[usenatbib]{mn2e}
\usepackage{apjfonts}
\usepackage{amssymb}
\usepackage{amsmath}
\usepackage{ctable}
\usepackage{fixltx2e} 

\newcommand{\be}{\begin{equation}}
\newcommand{\ee}{\end{equation}}

\newcommand{\rhoX}{\rho}
\newcommand{\lnrhoX}{\ln{\rho}}
\newcommand{\uv}{\tilde{v}}

\newcommand\plotonesize[2]
 {\centering \leavevmode \includegraphics[width={#2\columnwidth}]{#1}}

\newcommand{\acknowledgments}{\begin{small}\section*{Acknowledgments}\end{small}}
\newcommand\altaffilmark[1]{$^{#1}$}
\newcommand\altaffiltext[1]{$^{#1}$}
\title[Intermittent Density PDFs]{A Model for (Non-Lognormal) Density Distributions in Isothermal Turbulence
\vspace{-0.5cm}}

\author[Hopkins]{
\parbox[t]{\textwidth}{ 
Philip F. Hopkins\altaffilmark{1}\thanks{E-mail:phopkins@astro.berkeley.edu}} 
\vspace*{6pt} \\
\altaffiltext{1}{Department of Astronomy, University of California
  Berkeley, Berkeley, CA 94720\vspace{-1.1cm}} \\
}

\date{Submitted to MNRAS, September, 2012\vspace{-0.6cm}}
\begin{document}
\maketitle
\label{firstpage}

\begin{abstract}
\vspace{-0.2cm}

We propose a new, physically motivated fitting function for density PDFs in turbulent, ideal gas. Although it is generally known that when gas is isothermal, the PDF is approximately lognormal in the core, high-resolution simulations show large deviations from exact log-normality. The proposed function provides an extraordinarily accurate description of the density PDFs in simulations with Mach numbers $\sim0.1-15$ and dispersion in $\log{(\rho)}$ from $\sim0.01-4$\,dex. Compared to a lognormal or lognormal-skew-kurtosis model, the fits are improved by orders of magnitude in the wings of the distribution (with fewer free parameters). This is true in simulations using a variety of distinct numerical methods, including or excluding magnetic fields. Deviations from lognormality are represented by a parameter $T$ that appears to increase systematically with the compressive Mach number of the simulations. The proposed distribution can be derived from intermittent cascade models of the longitudinal (compressive) velocity differences, which should be directly related to density fluctuations, and we also provide a simple interpretation of the density PDF as the product of a continuous-time relaxation process. As such this parameter $T$ is consistent with the same single parameter needed to explain the (intermittent) velocity structure functions; its behavior is consistent with turbulence becoming more intermittent as it becomes more dominated by strong shocks. It provides a new and unique probe of the role of intermittency in the density (not just velocity) structure of turbulence. We show that this naturally explains some apparent contradictory results in the literature (for example, in the dispersion-Mach number relation) based on use of different moments of the density PDF, as well as differences based on whether volume-weighted or mass-weighted quantities are measured. We show how these are fundamentally related to the fact that mass conservation requires violations of log-normal statistics.

\end{abstract}

\begin{keywords}
hydrodynamics --- instabilities --- turbulence --- star formation: general --- galaxies: formation --- galaxies: evolution --- galaxies: active --- cosmology: theory
\vspace{-1.0cm}
\end{keywords}

\vspace{-1.1cm}
\section{Introduction}
\label{sec:intro}

It is well-established that idealized isothermal turbulence (i.e.\ driven turbulence in the absence of external forces) drives the density probability distribution function (PDF) to an approximately log-normal shape (a Gaussian in $\ln{\rho}$), with a dispersion that increases weakly with the Mach number \citep[see e.g.][]{vazquez-semadeni:1994.turb.density.pdf,padoan:1997.density.pdf,scalo:1998.turb.density.pdf,ostriker:1999.density.pdf,klessen:2000.pdf.supersonic.turb}. This is easily understood: the Navier-Stokes equations for inviscid, isothermal gas can be written only in terms of $\ln{\rho}$ (i.e.\ changes due to the velocity field are multiplicative), so if the density distribution is the product of a sufficient number of random, uncorrelated and unbiased multiplicative perturbations it should converge via the central limit theorem to a lognormal \citep{passot:1998.density.pdf,nordlund:1999.density.pdf.supersonic}. This assumption has become widespread, and forms the basis for interpretation of many observations \citep[e.g.][]{ossenkopf:2002.obs.gmc.turb.struct,ridge:2006.lognormal.pdf,wong:2008.gmc.column.dist}, as well as a wide range of theoretical models for ISM structure and star formation (\citealt{elmegreen:2002.fractal.cloud.mf,padoan:2002.density.pdf,krumholz.schmidt,hennebelle:2008.imf.presschechter,hopkins:excursion.clustering,hopkins:excursion.ism,hopkins:excursion.imf,hopkins:excursion.imf.variation,federrath:2012.sfr.vs.model.turb.boxes,federrath:2012.sfe.pwrspec.vs.time.sims}).

\begin{figure}
    \centering
    \plotonesize{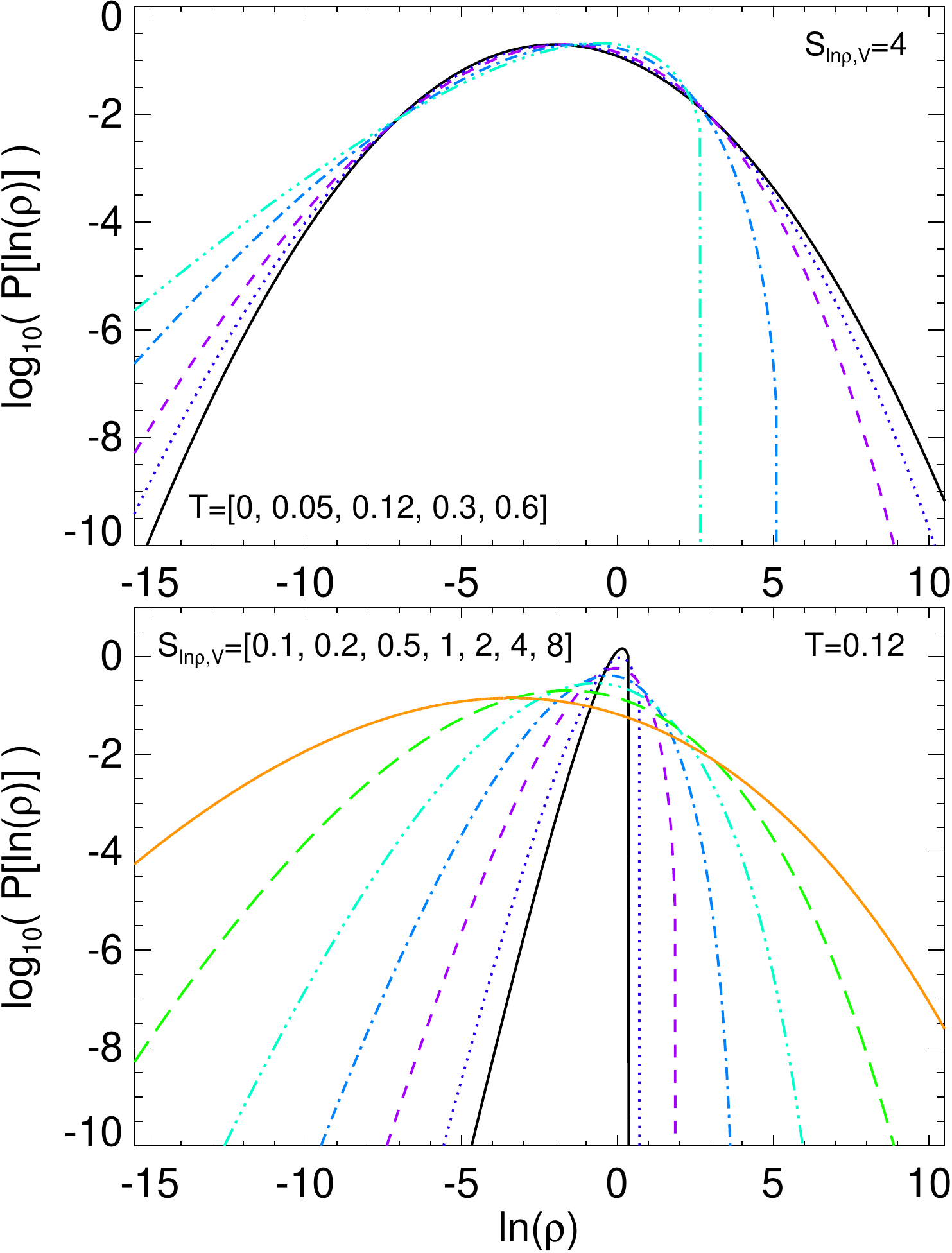}{0.85}
    \caption{Examples of the proposed (volume-weighted) log-density PDF (Eq.~\ref{eqn:PV}), as a function of parameters $T$ (increasing with deviations from log-normal) and $S_{\ln{\rho},\,V}$ (volume-weighted variance in $\ln{\rho}$). {\em Top:} Varying $T$ at fixed variance. $T=0$ is lognormal (the black solid line); the skew increases systematically with $T$. {\em Bottom:} Changing variance at fixed $T$ (broader PDFs at larger $S_{\ln{\rho},\,V}$). A simulated box averaged over progressively larger scales should exhibit approximately fixed $T$ with decreasing $S_{\ln{\rho},\,V}$, varying the PDF shape as shown.
    \label{fig:demo.pdf}}
\end{figure}

\begin{figure}
    \centering
    \plotonesize{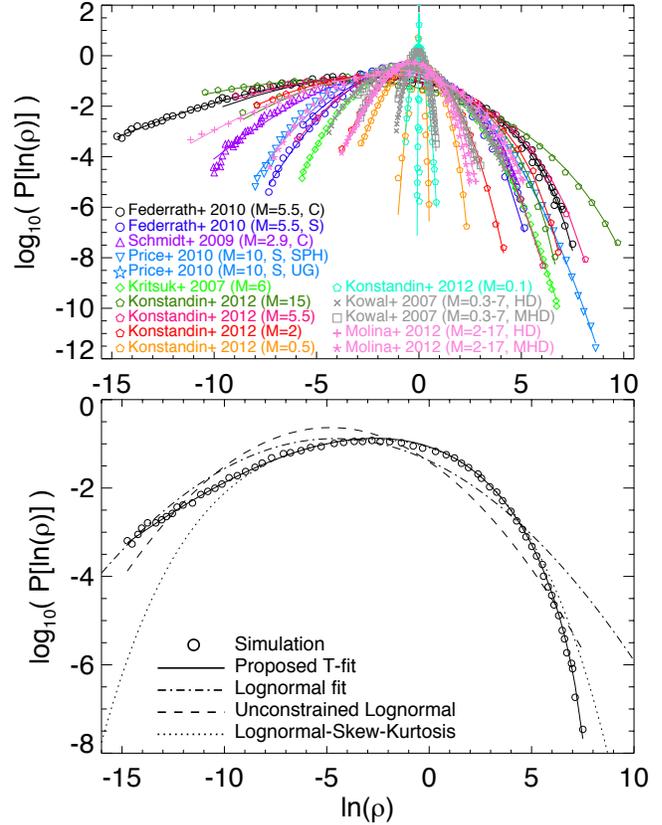}{1.02}
    \caption{Fits to simulated isothermal turbulent boxes. {\em Top:} Volume weighted PDFs from simulations (points) in Table~\ref{tbl:sims}, spanning a range in Mach numbers and variances, driving schemes, numerical methods, and magnetic field strengths. In the same color, lines show the best-fit with the proposed Eq.~\ref{eqn:PV}. {\em Bottom:} One example (the \citealt{federrath:2010.obs.vs.sim.turb.compare} compressively-forced $\mathcal{M}=5.6$ model), comparing the best-fit with Eq.~\ref{eqn:PV} to a standard lognormal fit, a lognormal fit with independently freed normalization and mean (i.e.\ not constrained to be properly normalized or conserve mass), and a lognormal-skew-kurtosis model (fourth order log-log polynomial). The proposed model fits extremely well, whereas even higher-order lognormal models (with more free parameters) fare very poorly (especially in the tails of the PDF). 
    \label{fig:fit.pdf}}
\end{figure}

However, a growing number of higher-resolution simulations show deviations from exactly log-normal behavior in driven isothermal turbulence, which can become very large (orders-of-magnitude) in the tails of the distributions \citep[see e.g.][]{federrath:2008.density.pdf.vs.forcingtype,federrath:2010.obs.vs.sim.turb.compare,federrath:2012.sfe.pwrspec.vs.time.sims,schmidt:2009.isothermal.turb,price:2010.grid.sph.compare.turbulence,konstantin:mach.compressive.relation}. This appears to be especially severe when the density PDFs are broad (high-variance). And there are well-known, highly statistically significant inconsistencies between the properties inferred from the volume-weighted versus mass-weighted density PDFs, or between different moments of the density PDFs, which should not appear if they were truly log-normal \citep[see above and][]{lemaster:2009.density.pdf.turb.review,burkhart:2009.mhd.turb.density.stats,federrath:2010.obs.vs.sim.turb.compare,price:2011.density.mach.vs.forcing,molina:2012.mhd.mach.dispersion.relation}.

In fact, mass conservation fundamentally invalidates the assumptions that lead to the log-normal prediction. First, consider the density distribution sampled within a large box, measured as the volume-average density $\rho_{i}=m_{i}/v_{0}$ in independent, equal, small sub-volumes $v_{0}$, and assume it is exactly lognormal. The density distribution smoothed on a larger scale $v_{1}\gg v_{0}$ is then necessarily the sum over the sub-volumes $v_{0,\,i}$ inside $v_{1}$, $\rho_{j} = (\sum m_{i})/v_{1} = (v_{0}/v_{1})\,\sum \rho_{i}$, i.e.\ it is the {\em linear} sum of lognormally distributed variables. But the convolution of linear values of lognormal variates is not lognormal (although it can be approximately so; see \citealt{fenton:1960.lognormal.sum.approx}). So mass conservation implies the density distribution cannot be lognormal on more than one scale. Second, convergence to a lognormal PDF via the central limit theorem requires fluctuations be uncorrelated; but this cannot be true in detail. Consider again a small but {\em finite} volume $v_{0}$ within some larger volume containing total mass $M$; there is nothing stopping there from being arbitrarily small mass within $v_{0}$ (i.e.\ fluctuations can extend to infinitely large in negative $\ln{\rho}$), but mass conservation sets an upper limit to the maximum {\em volume-averaged} density within the volume ($\ln{\rho}\le \ln{(M/v_{0})}$; the local density on smaller scales is free to diverge). So especially for large fluctuations, which may be important in the tails of the distribution, mass conservation implies a biased/asymmetric distribution in $\ln{\rho}$ sampled at finite resolution. There are many other reasons the PDF may be non-lognormal; for example, it is well-known that compressive (longitudinal) velocity fluctuations are not exactly Gaussian-distributed, owing to intermittency (discrete phenomena such as shocks or waves producing extended tails in the PDF), and there are (small) residual correlations between the local Mach number and density which imply correlated fluctuations \citep[see][]{passot:1998.density.pdf,kritsuk:2007.isothermal.turb.stats,federrath:2010.obs.vs.sim.turb.compare}.

This is important for several reasons. Many of the physically interesting behaviors in turbulence, and models described above, rely critically on the behavior in the extremes of the distribution (where deviations might be very large). Also, the assumption that PDFs are lognormal, and hence that their moments are related by certain simple analytic scalings, has led to widespread use of ``proxies'' for other quantities of interest (for example, it is common to infer, rather than measure, the dispersion in $\ln{(\rho)}$, by actually measuring the mean value, and assuming such a linking relation) -- these could be seriously in error. This could severely bias estimates of the dispersion-Mach number relation in different studies. Non-normal behavior in density PDFs may provide a valuable additional probe of intermittency in compressible turbulence, which remains poorly understood. And the differences from lognormal behavior, especially in how the volume-weighted and mass-weighted PDFs relate, are critical to relate results from different simulations to one another and to understand convergence in different numerical methods.

In this paper, we propose a new functional form for the density distribution in isothermal, driven turbulence. We show that this provides an excellent fit to the density PDF in simulations (significantly improved over a lognormal model), over a wide dynamic range. We also show that this proposed PDF resolves the inconsistencies with mass conservation discussed above, and can be derived from models of intermittency in turbulent velocity fluctuations, linking the non-normal features in density PDFs to fundamental properties of compressible turbulence.

\begin{figure}
    \centering
    \plotonesize{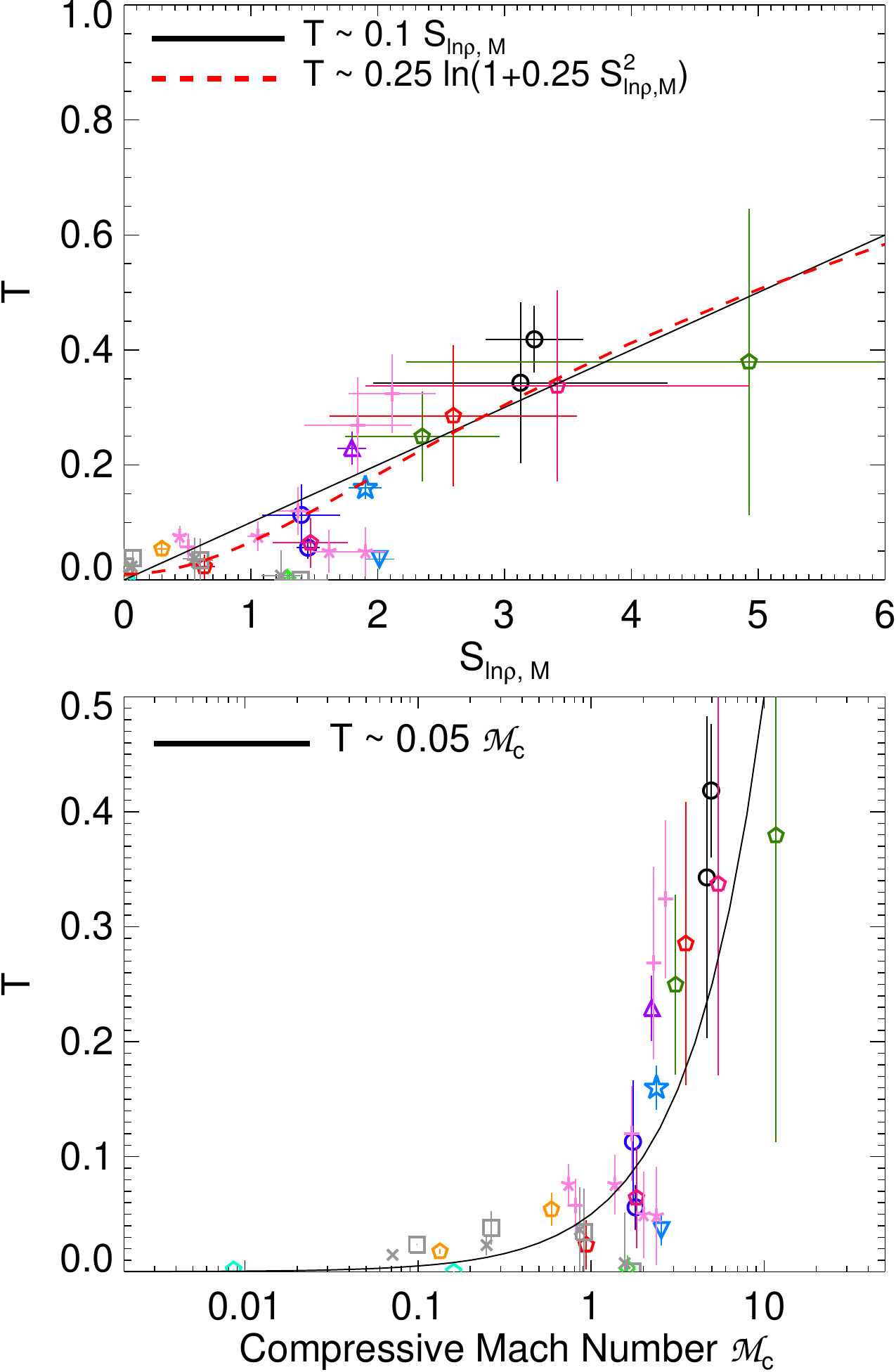}{0.85}
    \caption{{\em Top:} Fitted parameter $T$ representing the deviations from lognormal, as a function of mass-weighted variance in $\ln{\rho}$, for each simulation in Table~\ref{tbl:sims}. $T$ appears to increase with $S_{\ln{\rho},\,M}$, in simple fashion. {\em Bottom:} Same, but versus the mean compressive component of the driving Mach number $\mathcal{M}_{c}$ ($\mathcal{M}_{c}\approx\mathcal{M}$ for pure compressive driving, $\approx \mathcal{M}/3$ for solenoidal driving). The dependence is even more pronounced, with $T$ rising sharply at $\mathcal{M}_{c}\gtrsim1$, approximately as $T\approx0.05\,\mathcal{M}_{c}$.
    \label{fig:T.vs.S}}
\end{figure}

\begin{figure}
    \centering
    \plotonesize{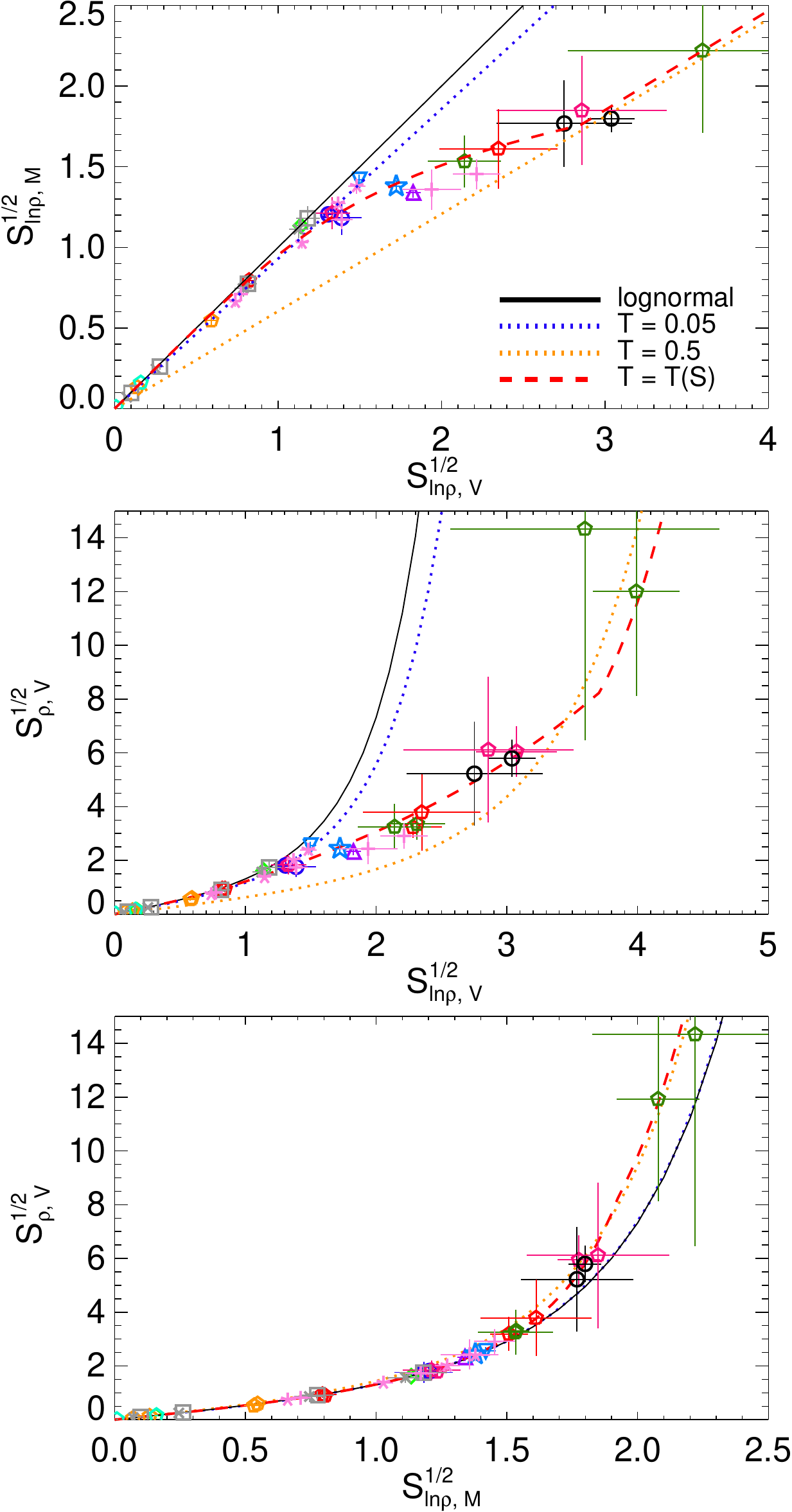}{0.95}
    \caption{Moments of the density distribution. The simulations from Table~\ref{tbl:sims} are shown, with the prediction for a lognormal model (Eqs.~\ref{eqn:LN.moment.V}-\ref{eqn:LN.moment.M}), and our proposed Eq.~\ref{eqn:PV} (see Eqs.~\ref{eqn:moment.V}-\ref{eqn:moment.M}) for fixed $T=0.05$, $T=0.5$ and the best fit $T = T(S) \propto \mathcal{M}_{c}$ from Fig.~\ref{fig:T.vs.S}. 
    {\em Top:} Mass-weighted dispersion in $\ln{\rho}$ ($=S_{\ln{\rho},\,M}^{1/2}$) versus volume-weighted dispersion in $\ln{\rho}$ ($=S_{\ln{\rho},\,V}^{1/2}$). These are equal in a lognormal model. The simulations do not follow this relation, but trace the mean $T=T(S)$ track between $T=0.05$ and $T=0.5$.
    {\em Middle} Volume-weighted dispersion in the linear $\rho$, versus volume-weighted dispersion in $\ln{\rho}$. The simulations lie well below the lognormal prediction. Deviations from lognormal are significant at {\em all} $S$ here (though difficult to see at smaller values), but agree well with the proposed model.
    {\em Bottom:} Volume-weighted dispersion in the linear $\rho$, versus mass-weighted dispersion in $\ln{\rho}$. Although the data still favor the non-lognormal model, it is worth noting that deviations from the lognormal case are minimized when comparing these specific moments. 
    \label{fig:S.moments}}
\end{figure}

\vspace{-0.5cm}
\section{Models for the Density PDF}

Consider the density PDF. We define first the {\em volume-weighted} PDF of log-density,  
$P_{V}(\lnrhoX)$, i.e.\ the probability of a given unit volume $V$ having an averaged density $\rho = M(V)/V$, per unit $\lnrhoX$.
This must obey two constraints:
\begin{align}
\label{eqn:constraints}
\int_{-\infty}^{\infty}P_{V}(\lnrhoX)\,{\rm d}\lnrhoX &= 1 \\ 
\nonumber \int_{-\infty}^{\infty}\,\rhoX\,P_{V}(\lnrhoX)\,{\rm d}\lnrhoX &= \rho_{0} \equiv 1 
\end{align}
the first condition is just that the PDF be properly normalized, the second is mass conservation (integrated over all volumes $V$, the total mass $M$ is conserved: $\int\,\rho\,V_{\rm tot}\,P_{V}\,{\rm d}\lnrhoX = M = \rho_{0}\,V_{\rm tot}$). Throughout this paper, we use dimensionless units such that the volume averaged mean density $\rho_{0}=1$.

We can also define the {\em mass-weighted} density PDF, $P_{M}(\lnrhoX)$. Since $\rho\equiv {\rm d}M/{\rm d}V$, it follows that $P_{M}(\lnrhoX)=\rhoX\,P_{V}(\lnrhoX)$ \citep[see e.g.][]{li:2003.turb.mc.vs.eos}. 

\vspace{-0.5cm}
\subsection{Lognormal PDFs}

In the simple log-normal case, we have:
\be
\label{eqn:PV.LN}
P_{V}^{\rm LN}(\lnrhoX) = \frac{1}{\sqrt{2\pi\,S_{\ln{\rho},\,V}}}\,\exp{{\Bigl[}-\frac{(\lnrhoX+S_{\ln{\rho},\,V}/2)^{2}}{2\,S_{\ln{\rho},\,V}}{\Bigr]}}
\ee
where $S_{\ln{\rho},\,V}$ is the volume-weighted variance in $\lnrhoX$ ($S_{\ln{\rho},\,V}\equiv \int (\lnrhoX-\langle\lnrhoX\rangle)^{2}\,P_{V}(\lnrhoX)\,{\rm d}\lnrhoX$). The normalization and mean are fixed by Eq.~\ref{eqn:constraints}. 
In this case, the various volume-weighted moments of $\rhoX$ and $\lnrhoX$ are easily calculated: 
\begin{align}
\label{eqn:LN.moment.V}
\langle\rhoX\rangle_{V}^{\rm LN} &= 1 \\ 
\nonumber \langle\lnrhoX\rangle_{V}^{\rm LN} &= -S_{\ln{\rho},\,V}/2 \\ 
\nonumber S_{\rho,\,V}^{\rm LN} &=\exp{(S_{\ln{\rho},\,V})} - 1
\end{align}
The mass-weighted PDF $P_{M}(\lnrhoX)$ is also lognormal, with moments
\begin{align}
\label{eqn:LN.moment.M}
S_{\ln{\rho},\,M}^{\rm LN} &= S_{\ln{\rho},\,V} \\ 
\nonumber \langle\lnrhoX\rangle_{M}^{\rm LN} &= +S_{\ln{\rho},\,M}/2 \\ 
\nonumber \langle\rhoX\rangle_{M}^{\rm LN} &= \exp{(S_{\ln{\rho},\,M})} \\ 
\nonumber S_{{\rho},\,M}^{\rm LN} &= \exp{(3\,S_{\ln{\rho},\,M})} - \exp{(2\,S_{\ln{\rho},\,M})} 
\end{align}

\vspace{-0.5cm}
\subsection{An Improved Model}

Motivated by the considerations in \S~\ref{sec:intermittency} \&\ Appendix~\ref{sec:appendix:deriv}, consider the following simple model for the volumetric density PDF: 
\begin{align}
\label{eqn:PV} 
P_{V}(\ln{\rho})\,{\rm d}\ln{\rho} 
&= I_{1}(2\,\sqrt{\lambda\,u})\,\exp{[-(\lambda+u)]}\,\sqrt{\frac{\lambda}{u}}\,{\rm d}u \\ 
\nonumber &= \sum_{m=0}^{\infty}\,\frac{\lambda^{m}\,e^{-\lambda}}{m!}\,\frac{u^{m-1}\,e^{-u}}{(m-1)!}\,{\rm d}u \\ 
\nonumber u &\equiv \frac{\lambda}{1+T} -  \frac{\ln{\rho}}{T}\ \ \ \ \ \ \ \ \ \ (u\ge0) \\ 
\nonumber \lambda &\equiv \frac{S_{\ln\rho,\,V}}{2\,T^{2}} 
\end{align}
where $I_{1}(x)$ is the modified Bessel function of the first kind,\footnote{$I_{1}(x)$ satisfies
\begin{align}
\nonumber I_{1}(x) &= \sum_{m=0}^{\infty}\,\frac{(x/2)^{2\,m+1}}{m!\,(m+1)!} \\ 
\nonumber x^{2}\,\frac{\partial^{2}I_{1}}{\partial x^{2}} &+ x\,\frac{\partial I_{1}}{\partial x} - (x^{2}+1)\,I_{1} = 0
\end{align}
A useful relation for large $x\gtrsim100$ is 
\be
\nonumber I_{1}(x) \rightarrow (2\pi\,x)^{-1/2}\,\exp{(x)}
\ee 
}
and $P_{V} = 0$ for $u<0$. 
It is straightforward to verify that this obeys both constraints in Eq.~\ref{eqn:constraints}. Given some variance $S_{\ln{\rho},\,V}$ (or equivalently, parameter $\lambda$), then, this is a one-parameter model in the value of the adjustable parameter $T$. We will discuss the meaning of this parameter below.

From this, we can immediately derive the following useful relations for the moments:
\begin{align}
\label{eqn:moment.V}
S_{\ln{\rho},\,V} &= 2\,\lambda\,T^{2} \\ 
\nonumber S_{\rho,\,V} &= \exp{{\Bigl(}\frac{2\,\lambda\,T^{2}}{1+3\,T+2\,T^{2}}{\Bigr)}} - 1 \\
\nonumber &= \exp{{\Bigl(}\frac{S_{\ln{\rho},\,V}}{1+3\,T+2\,T^{2}}{\Bigr)}} - 1 \\
\nonumber \langle \ln{\rho} \rangle_{V} &= -\frac{\lambda\,T^{2}}{1+T} = -\frac{S_{\ln{\rho},\,V}}{2}\,(1+T)^{-1} \\
\nonumber \langle \rho \rangle_{V} &= 1 \\ 
\label{eqn:moment.M}
S_{\ln{\rho},\,M} &= \frac{2\,\lambda\,T^{2}}{(1+T)^{3}} = {S_{\ln{\rho},\,V}}\,{(1+T)^{-3}} \\ 
\nonumber S_{\rho,\,M} &= \exp{{\Bigl(}\frac{6\,\lambda\,T^{2}}{1+4\,T+3\,T^{2}}{\Bigr)}} - 
\exp{{\Bigl(}\frac{4\,\lambda\,T^{2}}{1+3\,T+2\,T^{2}}{\Bigr)}} \\
\nonumber &= \exp{{\Bigl(}\frac{3\,S_{\ln{\rho},\,M}\,(1+T)^{3}}{1+4\,T+3\,T^{2}}{\Bigr)}} - 
\exp{{\Bigl(}\frac{2\,S_{\ln{\rho},\,M}\,(1+T)^{3}}{1+3\,T+2\,T^{2}}{\Bigr)}}  \\
\nonumber \langle \ln{\rho} \rangle_{M} &= \frac{\lambda\,T^{2}}{(1+T)^{2}} = +\frac{S_{\ln{\rho},\,M}}{2}\,(1+T) \\ 
\nonumber \langle \rho \rangle_{M} &= \exp{{\Bigl(}\frac{2\,\lambda\,T^{2}}{1+3\,T+2\,T^{2}}{\Bigr)}}
= \exp{{\Bigl(}\frac{S_{\ln{\rho},\,M}\,(1+T)^{3}}{1+3\,T+2\,T^{2}}{\Bigr)}}
\end{align}

The quantity $T$ clearly represents a degree of non-lognormality, and breaks the similarity otherwise seen between e.g.\ the volume-weighted and mass-weighted distributions. Note that as $T\rightarrow0$, these relations do simplify to those of the lognormal distribution; Eq.~\ref{eqn:PV} becomes identical to Eq.~\ref{eqn:PV.LN}.
Fig.~\ref{fig:demo.pdf} illustrates the shape of this PDF for different $T$ and $S_{\ln{\rho},\,V}$.

\vspace{-0.5cm}
\section{Results}
\label{sec:results}

Fig.~\ref{fig:fit.pdf} compares our proposed Eq.~\ref{eqn:PV} to an ensemble of simulation results compiled form the literature. Each simulation is an idealized driven ``turbulent box,'' but we intentionally consider a wide range of results using different turbulent forcing schemes (solenoidal vs.\ compressive), strength/Mach numbers (from $\mathcal{M}\sim0.1-15$), numerical methods (fixed grid, adaptive-mesh refinement, and smooth-particle hydrodynamics), and resolution. The properties of each simulation are summarized in Table~\ref{tbl:sims}. For our purposes this allows us to examine the PDF over a wide dynamic range. In each case we take the volume-weighted PDF $P_{V}$ (or convert $P_{M}$ presented by the authors to $P_{V}$). 

We fit each simulated PDF taking $S_{\ln{\rho},\,V}$ as fixed to the {\em true} variance in each distribution,\footnote{We define the ``true'' variance as the variance measured directly from the simulation density PDF data points (either mass or volume-weighted as appropriate), without reference to any fitting function.} so that there is only one free parameter $T$. In every case, our proposed function fits the data exceptionally well; there is no case with $\chi^{2}/\nu\gg1$.\footnote{The fits are also remarkably stable. If, for example, we also free $S_{\ln{\rho},\,V}$, we recover nearly identical answers (well within the $1\sigma$ error bars), albeit now with two free parameters. We can in principle also free the normalization and value of $\rho_{0}$ (though this would violate the constraint Eq.~\ref{eqn:constraints}) and recover the same answer within $\sim1\%$. We have also experimented with randomly removing $1,\,10,\,30,\,50$ and $75\%$ of the data points in every PDF, and find in every case that the results remain consistent with their $1\sigma$ uncertainties.} In contrast, in a majority of cases (and every case with $S_{\ln{\rho},\,V}\gtrsim1$) the lognormal model is inaccurate ($\chi^{2}/\nu\gg1$). 

This is illustrated for one typical PDF, with a large variance $S_{\ln{\rho},\,V}\approx9.2$. We show the simulation data and our best-fit with $T\approx0.4$. The RMS deviation $\delta_{\rm rms}$ of the simulation data about the best-fit is $<0.05\,$dex. In contrast, the lognormal model with the correct variance $S_{\ln{\rho},\,V}$ has $\delta_{\rm rms}\approx 0.77$\,dex. If we free all parameters in the lognormal model, so that it now has three degrees of freedom (as opposed to one in our proposed model), the best-fit $\delta_{\rm rms}=0.51$\,dex (moreover, this model produces a PDF which is non-unitary by a factor of $\sim5$, fails at mass conservation by a factor of $\approx80$, and under-predicts the correct variance by $\sim1.2\,$dex). We could simply consider higher-order polynomials in log-log space: for example, fitting a lognormal-skew-kurtosis model (i.e.\ adding third and fourth-order terms in the exponential as proposed in \citet{federrath:2010.obs.vs.sim.turb.compare}); but even with these added degrees of freedom we find the best-fit has $\delta_{\rm rms}=0.4$\,dex. In fact, provided that the necessary physical constraints (Eq.~\ref{eqn:constraints}) are obeyed, we require an {\em eighth}-order log-log polynomial to recover the same accuracy in $\delta_{\rm rms}$ as our proposed fit. This not only introduces seven additional free parameters, but it also is extremely unstable (the results can change by orders of magnitude if we add/remove points in the wings of the distribution). 

Fig.~\ref{fig:T.vs.S} shows the magnitude of $T$ (the strength of deviations from lognormal statistics) fit to each simulation, as a function of variance. There is a clear trend, where the deviations are small (although we emphasize in many cases still extremely statistically significant) at small $S$ then rise with increasing $S$. We discuss possible reasons for this below, but for now simply note it can be parameterized by a simple linear fit $T\sim 0.1\,S_{\lnrhoX,\,M}$, or, more accurately, as proportional to the compressive component of the Mach number $\mathcal{M}_{c}$ (see Table~\ref{tbl:sims}), $T\sim 0.05\,\mathcal{M}_{c}$. 

Fig.~\ref{fig:S.moments} shows how the proposed model can explain many deviations from lognormal statistics seen in the moments of these distributions. First, we compare the standard deviations ($=S^{1/2}$) of $\lnrhoX$ in both mass and volume-weighted measures. If the PDFs were lognormal, $S_{\lnrhoX,\,M}=S_{\lnrhoX,\,V}$; but we see this clearly is not the case (the lognormal prediction is ruled out as the null hypothesis here at $>6\,\sigma$ confidence). Simulations always find $S_{\lnrhoX,\,M}< S_{\lnrhoX,\,V}$, with the differences being especially large at high variance. We compare the predicted relations for $T=0.05$ and $T=0.5$, which bracket the data at low and high variance, respectively. We also compare the track for the mean linear relation $T\propto \mathcal{M}_{c}$, which runs through the locus of points. 

We next compare the (volume-weighted) dispersion in the linear density ($S_{\rho,\,V}^{1/2}$) to that in the logarithmic density. A lognormal model predicts $S_{\rho,\,V}=\exp{(S_{\ln{\rho},\,V})}-1$; this is again clearly ruled out ($>8\,\sigma$ confidence). Again simulations lie between the predicted relations for $T=0.05$ and $T=0.5$, with the mean $T-S$ track predicting a curve that goes through essentially all points. Interestingly, if we compare $S_{\rho,\,V}^{1/2}$ to the {\em mass-weighted} dispersion in the log density $S_{\ln{\rho},\,M}$, it turns out that the deviations from the lognormal prediction with $T$ are minimized. The data still favor a model with $T>0$, but the predicted tracks differ by a small amount. This explains why certain properties appear to agree better with lognormal models when taken as mass-weighted (discussed below; see also \citealt{konstantin:mach.compressive.relation}).


\vspace{-0.5cm}
\section{A Physical Model for the Density PDF}
\label{sec:intermittency}

\subsection{Quantized Log-Poisson Cascades}
\label{sec:intermittency:steps}

We now consider a physical motivation for Eq.~\ref{eqn:PV}. Perhaps the most popular model for the statistics of intermittent turbulence is the \citet{sheleveque:structure.functions} (hereafter SL) cascade model. In strictly self-similar (Kolmogorov) turbulence, the velocity structure functions $S_{p}(R)=\langle \delta v(R)^{p} \rangle \equiv \langle |{\bf v}({\bf x}) - {\bf v}({\bf x}+{\bf R})|^{p} \rangle$ scale as power laws $S_{p}(R)\propto R^{\zeta_{p}}$ with $\zeta_{p}=p/3$. The SL model proposed an alternative phenomenological scaling:
\be
\label{eqn:sheleveque.structurefn}
\zeta_{p} = (1-\gamma)\,\frac{p}{3} + \frac{\gamma}{1-\beta}\,{\Bigl(}1-\beta^{p/3}{\Bigr)}
\ee
with the original choices $\beta=\gamma=2/3$ (giving $\zeta_{p} = p/9 + 2\,[1-(2/3)^{p/3}]$). 

\citet{shewaymire:logpoisson} and \citet{dubrulle:logpoisson} showed that the scaling in Eq.~\ref{eqn:sheleveque.structurefn} was the exact result of a class of quantized log-Poisson statistics. They assume extended self-similarity, i.e.\ $\delta v(R)^{p} \propto R^{p/3}\,\epsilon_{R}^{p/3}$ (where $\epsilon_{R}$ is the dissipation term between scales), and a general hierarchical symmetry between scales -- such that the statistics of $\epsilon_{R}$ can be encapsulated in a term $\pi_{R}$ such that $\epsilon_{R} \sim \pi_{R}\,\epsilon_{R}^{\infty}$ (where $\epsilon_{R}^{\infty}$ describes the scaling of the average properties of the most extreme/singular objects, e.g.\ shocks). Under these conditions, SL-like scalings are equivalent to the statement that $\pi_{R}$ obeys log-Poisson statistics of the form
\begin{align}
\label{eqn:logpoisson}
P(\pi_{R})\,{\rm d}\pi_{R} &= {\rm d}Y\,\sum_{m}\,P_{\lambda(R)}(m)\,G_{R}(Y,\,m),\ \ \ Y \equiv \frac{\ln{\pi_{R}}}{\ln{\beta}} 
\end{align}
with
\begin{align}
P_{\lambda}(m) &= \frac{\lambda^{m}}{m!}\,\exp{(-\lambda)}
\end{align}
and $G_{R}$ is {any} well-defined, infinitely divisible probability distribution function (physically depending on the driving and character of ``structures'' in the turbulence). This describes a general class of random multiplicative processes that obey certain basic symmetry properties.

Since $\delta v(R) \propto \epsilon_{R}^{1/3} \propto \pi_{R}^{1/3}$, $\ln{(\delta v/\langle \delta v\rangle)}=\ln{\pi^{1/3}}=(1/3)\ln{\pi}$ is a linear transformation and should obey the same statistics as $(1/3)\,\ln{\pi_{R}}$ \citep[see e.g.][]{shewaymire:logpoisson,dubrulle:logpoisson}. And since the basic derivations of the relation between the density and velocity power spectra are based on the (linear) density field being the product of compressive modes in the velocity field (e.g.\ the dispersion-Mach number relation), $\delta\rho\equiv \rho/\rho_{0}-1$ obeys the same statistics as $\delta v(R)$ (for small ``steps'' in scale). Thus the statistics of $\lnrhoX$ should have the same form as $(1/3)\,\ln{\pi_{R}}$; and from Eqn.~\ref{eqn:logpoisson}, the statistics of $(1/3)\,\ln{\pi_{R}}$ are identical to the statistics of $\ln{\pi_{R}}$ for a value of $\beta\rightarrow \beta^{1/3}$. Thus, we might naturally expect
\be
\label{eqn:Plnrho.vs.Pvel}
P(\lnrhoX) = \frac{1}{\ln{\beta_{\rho}}}\sum_{m=0}^{\infty}P_{\lambda(R)}(m)\,G_{R}{\Bigl(}\frac{\lnrhoX}{\ln\beta_{\rho}},\, m {\Bigr)}
\ee
with $\beta_{\rho} = \beta^{1/3}$ (see \citet{hopkins:frag.theory}).

In \citet{shewaymire:logpoisson}, they assume the simplest possible case, that $G_{R}(Y,\,m)\rightarrow \delta(Y-m)$ (a Dirac $\delta$-function), which corresponds to all driving ``events'' (shocks, rarefactions, vortices) which produce velocity or density changes being strictly quantized (i.e.\ producing an exactly identical multiplicative effect in every instance). This would predict $P(\lnrhoX) \rightarrow P_{\lambda}(m = \lnrhoX/\ln{\beta_{\rho}})$. In such a quantized log-Poisson process, the density on a small scale $R$ would be related to the mean density on the box scale $L$ by $\rho(R) = \beta_{\rho}^{m}\,(R/L)^{\gamma}\,\rho_{0}$ where $m$ is Poisson-distributed with integer values (per $P_{\lambda}(m)$). While this can (for the appropriate choice of $\beta$ and $\gamma$) give structure functions in good agreement with experiment, it is too simple a model for the density PDF because it predicts $\ln{(\rho)}$ itself is quantized in units of $\Delta\,\ln{\rho}=\ln{\beta_{\rho}}$, rather than being a continuous variable.

\vspace{-0.5cm}
\subsection{Continuous Models for Intermittency}
\label{sec:appendix:alt.intermittency}

A more general statistical cascade model which produces continuous distributions is the ``thermodynamic'' model proposed in \citet{castaing:1996.thermodynamic.turb.cascade}. The initial assumptions there are quite different from SL (for example, the scaling of the ``most singular structure'' does not need to be assumed). However, as shown in \citet{hedubrelle:1998.thermo.equiv.logpoisson}, this is just a more general consistent limit of the same hierarchy (different choice of $G_{R}$). This model predicts directly the scaling of the longitudinal velocity increments ($\langle |\delta v|^{p} \rangle \propto \sigma_{r}^{p}$), rather than the indirectly related quantities $\pi_{R}$ or $\epsilon_{R}$, and models their scaling as:
\begin{align}
\label{eqn:thermo.var}
P(\uv)\,{\rm d}\uv &= \sum_{m=0}^{\infty}\,\frac{\lambda^{m}\,e^{-\lambda}}{m!}\,\frac{\uv^{m-1}\,e^{-\uv}}{(m-1)!}\,{\rm d}\uv 
\end{align}
where $\uv\equiv T_{v}^{-1}\,|\ln{(\sigma_{r}/\sigma_{L})}|$ with $T_{v}$ a constant.

We see immediately that this has the form of our proposed scaling for the form of $P_{V}(\lnrhoX)$ (with $T_{v}$ playing an identical role to $T$), with the assumption of Eq.~\ref{eqn:Plnrho.vs.Pvel} that we can make a one-to-one mapping between the statistics of the magnitudes in the log-velocity field and the log-density field -- i.e.\ that something like a dispersion-Mach number relation is universal across scales.

But it is also straightforward to see that this is just the general form of the log-Poisson statistics with 
\begin{align}
\label{eqn:GR.castaing}
G_{R}(Y\,|\,m)\,{\rm d}Y &\rightarrow \frac{\uv^{\,m-1}}{(m-1)!}\,\exp{(-\uv)}\,{\rm d}\uv \\ 
&=
\nonumber {\rm d}\uv \int {\rm d}^{n}\uv_{i}\,\delta{\Bigl(}\uv-\sum_{i}\uv_{i}{\Bigr)}\,\prod_{i=1}^{m}\,\exp{(-\uv_{i})}
\end{align}
The predicted structure functions are
\be
\label{eqn:structfn.castaing}
\zeta_{p} = \frac{p}{3}\,\frac{1+3\,T_{v}}{1+p\,T_{v}}
\ee
which -- although a nominally different form from Eq.~\ref{eqn:sheleveque.structurefn} -- gives nearly identical scalings for the choice $T_{v}\approx(\gamma/6)\,|\ln{\beta}|$ (at least to $p\sim20$).\footnote{For the ``standard'' SL assumptions in incompressible turbulence ($\beta=\gamma=2/3$),  $T_{v}\approx0.05$ gives nearly identical structure functions. For the \citet{boldyrev:2002.structfn.model} model often applied to super-sonic turbulence ($\beta=1/3$, $\gamma=2/3$), $T_{v}=0.12$ gives nearly identical structure functions.}

In \citet{castaing:1996.thermodynamic.turb.cascade}, the form above was motivated by basic considerations of the most general form of an infinitely divisible cascade. However, \citet{yakhot:deriv.thermo.structfn} independently obtained exactly the same scaling exponents (in the parameter $B\equiv1/T_{v}$ derived therein) studying directly the symmetry properties of the Navier-Stokes equations. Moreover they derive $T_{v}\approx1/20$, rather than adopting it phenomenologically (for the case of nearly-incompressible sub-sonic turbulence, appropriate to match the original SL model; $T_{v}$ should be larger for compressible turbulence). They also directly derive the PDF of compressible (longitudinal) velocity fluctuations, which is exactly what we require to derive the PDF of density fluctuations. And indeed to first order in $T_{v}\ll1$, the cascade model above captures the non-Gaussian behavior of the PDFs calculated in \citet{yakhot:deriv.thermo.structfn}.\footnote{To see this, insert the PDF of Eq.~\ref{eqn:thermo.var} into Eq.~29 of \citealt{yakhot:deriv.thermo.structfn}, dropping terms of $\mathcal{O}(T_{v}^{2})\sim 1/400$.}

For our purposes here, this cascade model of the density PDF also has a plain interpretation in terms of thinking of the density and velocity fields as the product of a continuous-time multiplicative random relaxation process. As with the quantized case, the variable $\lambda$ represents some ``number of events'' over a dynamic range $\ln{(r_{2}/r_{1})}$ in scale. But we see that the function $G_{R}$ in Eq.~\ref{eqn:GR.castaing} is the convolution over a Poisson waiting time distribution for each quantized number of events. In other words, the PDF in Eq.~\ref{eqn:PV} is a {stationary} result in a medium in which perturbations are driven by discrete, random events (point processes) with a constant average rate on a given scale, which then damp exponentially on a characteristic timescale (integrated over infinite timescales). $T_{v}^{-1}$ is a dimensionless ``rate parameter''; for small values of $T_{v}^{-1}$, events are widely spaced in time, so the variance between ``levels of decay'' at a given time is large, whereas for large values the variance is smaller (the medium is more smooth). The total variance of this distribution is $2\,\lambda\,T_{v}^{2}$, i.e.\ proportional to the ``number of events'' in some time, times the variance ``per event.'' 

\citet{chainais:2006.inf.divisible.cascade.review} show that such a compound log-Poisson cascade corresponds to the most general form of an infinitely divisible, multidimensional cascade which can be decomposed into a combination of point processes (``events'') and random multipliers (for example, the \citealt{kolmogorov:1962.lognormal.cascade} model, while ultimately producing some similar results, cannot be synthesized via a hierarchy of point processes except in one dimension).

\begin{figure}
    \centering
    \plotonesize{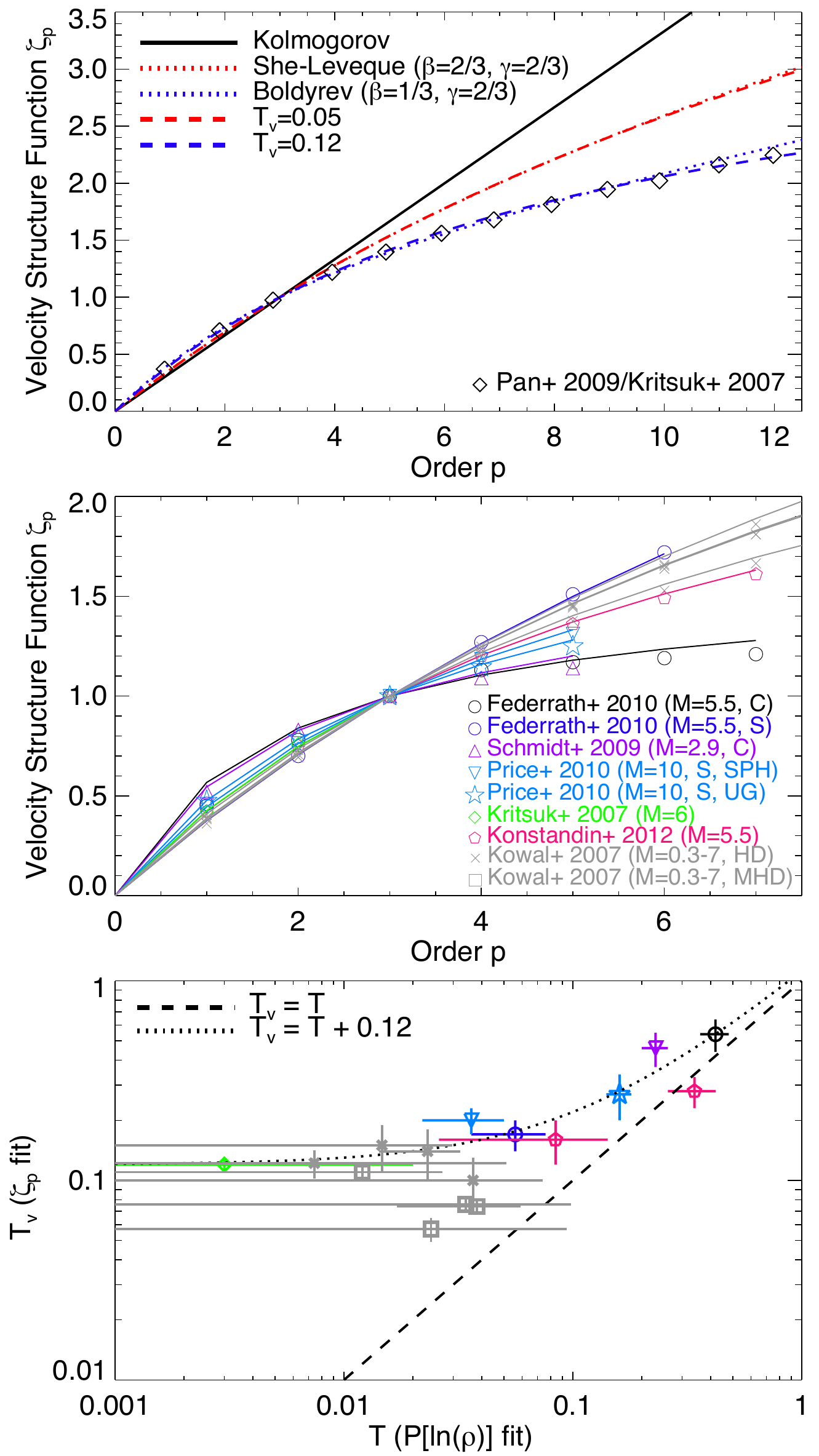}{0.93}
    \caption{Structure functions and their (potential) relation to non-lognormal density PDFs. 
    {\em Top:} Structure function scalings $\zeta_{p}$ (from $\langle \delta v(R)^{p} \rangle \propto R^{\zeta_{p}}$) versus order $p$. We compare the Kolmogorov prediction $\zeta_{p}=p/3$ (no intermittency), to the \citet{sheleveque:structure.functions} ``quantized log-Poisson'' model for intermittency (Eq.~\ref{eqn:sheleveque.structurefn}, with the labeled $\beta$, $\gamma$) and the \citet{boldyrev:2002.structfn.model} revision of this model for super-sonic turbulence. We also compare the prediction from \citet{castaing:1996.thermodynamic.turb.cascade} for the ``thermodynamic'' model (a continuous compound log-Poisson-exponential distribution), which predicts a PDF for longitudinal velocity moments that motivates our proposed density PDF, with a similar parameter $T_{v}$ that applies to the intermittency in the velocity field (Eq.~\ref{eqn:structfn.castaing}; for $T_{v}$ chosen to match the same assumptions as the other intermittency cases). The two produce nearly identical structure functions. Data from \citet{pan:boldyrev.structfn.tests} from one of the simulations we study (see Table~\ref{tbl:sims}) is shown.
    {\em Middle:} Structure functions $\zeta_{p}$ from the simulations in Fig.~\ref{fig:fit.pdf} (points), where available, and best-fit $T_{v}$ intermittency model.
    {\em Bottom:} Comparison of $T$ fit to the $\ln{\rho}$ distribution (Eq.~\ref{eqn:PV}) and $T_{v}$ fit to intermittency in the structure functions (Eq.~\ref{eqn:structfn.castaing}). The two increase together at large values, though there may be a ``floor'' in $T_{v}$ that is difficult to measure from the density PDF.
    \label{fig:structfn}}
\end{figure}

\begin{figure}
    \centering
    \plotonesize{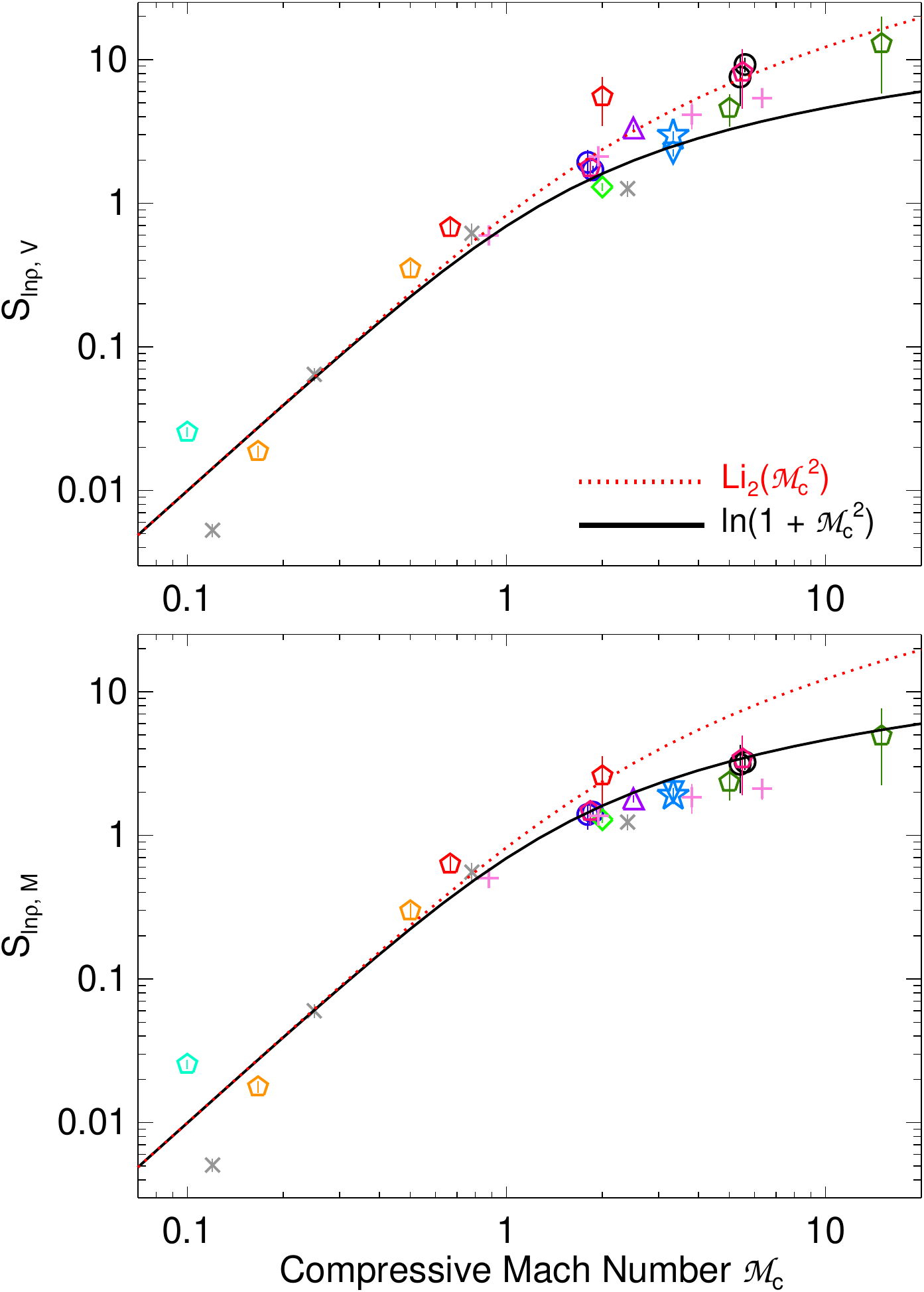}{0.95}
    \caption{Dispersion-Mach number relation, with the simulations from Table~\ref{tbl:sims} (only including those not magnetically-dominated). 
    {\em Top:} Volume-weighted dispersion in $\ln{\rho}$, versus compressive Mach number $\mathcal{M}_{c}$ (see Fig.~\ref{fig:T.vs.S}). Though a relation is clear, it does not follow the usual $S = \ln{(1+\mathcal{M}_{c}^{2})}$ curve (closer to Li$_{2}(\mathcal{M}_{c}^{2})$, the dilogarithm$^{\ref{foot:dilog}}$), and has large scatter at high-$\mathcal{M}_{c}$. The scatter becomes much larger if we just consider $\mathcal{M}$ instead of $\mathcal{M}_{c}$. 
    {\em Bottom:} Same, but using the mass-weighted dispersion in $\ln{\rho}$. Here the data agree very well with the predicted relation, with smaller scatter. This is related to the fact that the $S_{\ln{\rho},\,M}-S_{\rho,\,V}$ relation minimizes deviations from lognormal (i.e.\ allows for $S_{\rho,\,V}\sim \mathcal{M}_{c}^{2}$ with minimal corrections from non-lognormal terms). 
    \label{fig:S.Mach}}
\end{figure}

\vspace{-0.5cm}
\subsection{Testing this Explanation in Simulations}

We have already tested that the function in Eq.~\ref{eqn:PV} provides an exceptionally good fit to the simulation data. However we can further examine whether this might be connected with physical intermittency, as suggested by the model above. 

In Fig.~\ref{fig:structfn}, we show the velocity structure function scalings $\zeta_{n}$ predicted for pure self-similar turbulence, for the SL model with two commonly-adopted choices of $\beta$ and $\gamma$ (Eq.~\ref{eqn:sheleveque.structurefn}), and for the \citet{castaing:1996.thermodynamic.turb.cascade} model above with two values of $T_{v}$ (Eq.~\ref{eqn:structfn.castaing}). We denote in this section the value $T_{v}$ as value in Eq.~\ref{eqn:structfn.castaing} fit to the {\em velocity} structure functions, to clearly discriminate it from the $T$ in the fit to the density PDF. As noted above, for the appropriate choices of $T_{v}$, the SL and \citet{boldyrev:2002.structfn.model} structure functions are almost exactly reproduced, albeit with one fewer free parameter. We also compare the structure functions measured by \citet{pan:boldyrev.structfn.tests} from the simulations in \citet{kritsuk:2007.isothermal.turb.stats}, which agree very well with the \citet{boldyrev:2002.structfn.model} model as well as the $T_{v}=0.12$ \citet{castaing:1996.thermodynamic.turb.cascade} model (for another alternative intermittency model, see \citealt{schmidt:2008.turb.structure.fns}). Increasing $T_{v}$ represents systematically more intermittent turbulence, from $T_{v}\rightarrow0$ (Kolmogorov) to $T_{v}\rightarrow\infty$ ($\zeta_{n}\rightarrow1$ for all $n\ge 1$, as expected in pressure-free Burgers turbulence which is a random superposition of infinitely strong shocks). 

As noted above, the \citet{castaing:1996.thermodynamic.turb.cascade} $T_{v}$-models give nearly identical structure functions to the SL model for the choice 
\be
T_{v} \approx -\frac{\gamma}{6}\,\ln{\beta} = -\frac{\gamma}{6}\,\ln{{\Bigl(}1 - \frac{\gamma}{d-D_{\infty}}{\Bigr)}}
\ee
where here $d$ is the overall dimension of the system ($d=3$ in every case here) and $D_{\infty}$ is the ``fractal dimension'' of the ``most singular objects'' ($D_{\infty}=1$ for one-dimensional worms and lines, $D_{\infty}=2$ for sheet-like shock structures), while $\gamma$ is the singularity index (scaling of the size dependence of these structures; $\gamma=2/3$ for Kolmogorov, incompressible turbulence, while $\gamma=1$ for Burgers turbulence). 

We compile in Fig.~\ref{fig:structfn} the velocity structure functions measured for all the simulations for which we have measured density PDFs, wherever available (see Table~\ref{tbl:sims}). For each, we fit a model of the form of Eq.~\ref{eqn:structfn.castaing} to determine a best-fit $T_{v}$ (for convenience, assuming extended self-similarity; \citealt{benzi:1993.extended.selfsimilarity}). We see that there is a range of structure function shapes, but in every case a very good fit is possible with just one free parameter ($T_{v}$). 

Finally, we compare these $T_{v}$ values fitted to the velocity structure functions, to the $T$ determined from fitting the density PDF, for the same simulations. Unfortunately, the limited number of structure function determinations means this plot has little data at large $T$. However there are a couple of interesting notes. At $T\gtrsim 0.1$, we see $T_{v}\sim T$, albeit with just a few data points and non-trivial scatter. At small $T\lesssim 0.1$, we see that essentially all the velocity structure functions asymptote to values consistent with the prediction for \citet{boldyrev:2002.structfn.model}-like turbulence ($T_{v}\approx0.12$), even though the nominal $T$ fit to the density distributions can be much smaller. But we caution here that at very small $T$ the fractional uncertainties (fitting to the density PDF) are large (because the effects are very small), evident in the large error bars. Nevertheless there appears to be a trend in the sense that the simulations which have the largest deviations from log-normality in the density PDF also exhibit the largest deviation from even \citet{boldyrev:2002.structfn.model}-like scaling in the velocity structure functions -- both with the sense of being ``more intermittent.''

This is also consistent with the trend in Fig.~\ref{fig:T.vs.S} between $T$ and $S_{\ln{\rho}}$ or, similarly, compressive Mach number $\mathcal{M}_{c}$.

\vspace{-0.5cm}
\section{Discussion}
\label{sec:discussion}

We have proposed an alternative (non-lognormal), physically motivated fitting function to describe the density distributions in idealized simulations of isothermal turbulence. The proposed function, Eq.~\ref{eqn:PV}, provides a remarkably good fit to the data, over a huge range of Mach numbers and variance in the simulations, apparently independent of the driving mechanism (solenoidal vs.\ compressive), presence or absence of magnetic fields, and numerical methods in the simulations. 

Although it is widely assumed that density PDFs in isothermal turbulence are log-normal, we show that in fact a simple log-normal model is orders-of-magnitude less accurate as a description of these simulations, compared to the proposed function (especially in the wings of the distributions). Even higher order lognormal-skew-kurtosis models do little better. This manifests in a variety of ways, but particularly relevant is that the moments of the density distribution do not follow the predicted lognormal scalings (Eq.~\ref{eqn:LN.moment.V}-\ref{eqn:LN.moment.M}). 

This means, for example, that some previous studies which used the mean of $\ln{\rho}$ as a proxy for the variance, or assumed that the variance {should} be identical whether the PDF was volume-weighted or mass-weighted, could be significantly biased. The robustness of the fits presented here allows us to recover the variance $S$ in the density distribution nearly identically whether we measure it directly from the numerical PDF, or treat it as a fitting parameter. This is itself an important check of numerical convergence in simulations (i.e.\ it suggests that there is not sufficient mass/volume in unresolved tails of the distribution to significantly bias the dispersion-Mach number relation and other correlations). But the variance fitted in a lognormal model often differs from that directly calculated from the simulation data, and can depend significantly on how the fit is weighted (e.g.\ whether equally in linear or log $P(\ln{\rho})$).

Most simulations directly measure the variance used in the density dispersion-Mach number relation (without reference to a fitting function), but even in this case it is still important to specify how the statistics are weighted; in Fig.~\ref{fig:S.Mach} we show that quite different relations emerge depending on whether volume-weighted or mass-weighted dispersions are used in this correlation.\footnote{\label{foot:dilog}The mass-weighted variance $S_{\ln{\rho},\,M}$ appears to scale as $\ln{(1+\mathcal{M}_{c}^{2})}$, as generally adopted \citep[see][]{konstantin:mach.compressive.relation}. The volume-weighted variance $S_{\ln{\rho},\,V}$, however follows a steeper relation consistent with the trend we see in $T(S)$. Parameterized on its own, this appears to be closer to the dilogarithm Li$_{2}(\mathcal{M}_{c}^{2})$, where we can define 
\be
S_{\ln{\rho},\,V} \approx {\rm Li}_{2}(\mathcal{M}_{c}^{2}) \equiv \int_{0}^{1}\,\ln{(1+\mathcal{M}_{c}^{2}\,x)}\,
\frac{{\rm d}x}{x}
\ee
Interestingly, this is closer to the assumption that the total variance is the {\em sum} of the variance on all scales, where (scale-by-scale) the contribution to the variance is ${\rm d}S_{\ln{\rho},\,V}[R] \approx \ln{(1 + \langle \mathcal{M}_{c}^{2}[R] \rangle)}\,{\rm d}\ln{R}$, as proposed in \citet{hopkins:excursion.ism}.} This distinction is also critical to compare certain different numerical methods (Lagrangian vs.\ Eulerian codes). If it were neglected, one might arrive at different values of, for example, the $b$ parameter (relating $\mathcal{M}$ and $\mathcal{M}_{c}$) in the density dispersion-Mach number relation depending on the density weighting and numerical method. 

Even given the same variance, it is clear that the proposed fitting function with non-zero $T$ much more accurately describes the behavior of the ``wings'' of the density distribution. This may be especially important to models of star formation and fragmentation in the ISM (see references in \S~\ref{sec:intro}), which depend on calculating the mass or volume fraction above some threshold density (where e.g.\ self-gravity becomes important). If the threshold density of interest is sufficiently far outside the ``core'' of the distribution, our comparisons here suggest it may be critical to account for the non-lognormal character of the density PDFs. In extreme cases, orders-of-magnitude errors could result from simply extrapolating the density-Mach number relation and assuming a perfectly log-normal PDF (as is typically done in these models). 

We also provide a physical motivation for the proposed density PDF. We discuss (\S~\ref{sec:intro}) why mass conservation implies some asymmetry, breaking the conditions needed to drive the PDF to exactly lognormal; this asymmetry can be captured in the proposed PDF. And we specifically show that our proposed Eq.~\ref{eqn:PV} can be derived from multi-fractal cascade models of the longitudinal (compressive) velocity increments, in a more general formulation of the well-known \citet{sheleveque:structure.functions} hierarchy owing to \citet{castaing:1996.thermodynamic.turb.cascade}. This suggests that the strength of deviations from lognormal may be directly linked to intermittency, describable by the same parameter as the intermittent velocity structure functions. And it does appear that the same deviations from self-similarity appear directly in the structure functions of $\ln{\rho}$ \citep[see][]{kowal:2007.log.density.turb.spectra,liufang:2008.logpoisson.cosmic.baryons}. We also provide an alternative derivation based on a simple ``counts in boxes'' model in \S~\ref{sec:appendix:deriv}; this is heuristic (analogous to the discrete $\beta$ models used to describe some cases of the \citealt{sheleveque:structure.functions} model), but it is instructive and lends a physical interpretation to the parameters in the fit, based on the degree of variance allowed ``per step.'' In this scenario, the limit $T\rightarrow\infty$ corresponds to infinitely intermittent (Burgers) turbulence (a superposition of infinitely strong shocks), which would produce a highly non-normal density PDF, while $T\rightarrow0$ corresponds to a superposition of infinitely many smaller multiplicative perturbations, leading by the central limit theorem to a lognormal PDF. And indeed we appear to see in the simulations that as the strength of the compressive component of the Mach number increases -- i.e.\ as the fractional deviation induced by a strong shock, for example, becomes larger -- $T$ increases systematically. In practice, the largest values of $T$ we fit correspond to dimensions of intermittent structures $D_{\infty}\approx2.3$, which agrees well with what is directly measured in simulations of highly supersonic flows \citep{schmidt:2008.turb.structure.fns}. 

This provides an additional, powerful motivation to further study the deviations from log-normal behavior in density PDFs (independent of the pure accuracy of the fitting function). The $T$ parameter here  provides a direct diagnostic of the effects of intermittency on the density distribution in compressible turbulence, which remains poorly understood. To date, there has been very little analytic work developing a framework in which to understand intermittency in the density field (as opposed to the velocity field), though simulations probing this have advanced tremendously. Measuring $T$ alone, and understanding better how it scales with e.g.\ the Mach number of the simulations, can probe the nature of intermittent density structures and how they depend on the compressibility of turbulence. Already here we see hints of a continuous change in the character of these structures with $\mathcal{M}_{c}$. Comparing the $T$-values inferred from the density PDF to the full density and log-density structure functions provides a powerful and direct test of the cascade models described above. Related to this, understanding whether the density distribution does actually change shape with scale (in the manner predicted here), as is well-established for the velocity field, is fundamental to any full understanding of the statistics of compressible turbulence. And comparison of $T$ values and log-density structure functions to the velocity structure functions, expanding the limited statistics available here, can test whether the {\em same} structures dominate the statistics of both the density and velocity fields. In particular, this is critical to understand whether something like a density-Mach number relation can be generalized scale-by-scale.

That said, we strongly caution that some of the non-lognormal behavior in the PDFs fit here may be purely numerical. Because the high-density tail of the distribution is dominated by narrow, strong shocks and occupies little volume, the resolution demands are extreme, and convergence even at $>1000^{3}$ cells/particles may not be complete \citep[see][and references therein]{federrath:2010.obs.vs.sim.turb.compare}. Technical details of the forcing scheme may bias the low-density tail as well \citep{konstantin:mach.compressive.relation}. And differences remain between different numerical methods \citep{price:2010.grid.sph.compare.turbulence}; it is interesting to note that in the one otherwise identical experiment we compile, SPH methods appear to show smaller $T$ than fixed-grid methods (i.e.\ are more biased to high versus low densities), which may fundamentally relate to the Lagrangian vs.\ Eulerian nature of the codes. So considerable caution is needed when giving any physical interpretation to the fitted $T$, especially for any single simulation. That said, significant deviations from lognormal behavior manifest here even where the PDFs appear to be numerically converged. And in any case, as a pure phenomenological fitting function (independent of interpretation) the proposed function is a far more accurate description of the simulations than a lognormal.

Finally, we also caution that in real astrophysical turbulence, much larger deviations from the ideal behavior discussed here should be expected for many reasons. Real systems are not exactly isothermal, and this leads to strongly non-normal density PDFs (see e.g.\ \citealt{scalo:1998.turb.density.pdf}; though in Appendix~\ref{sec:appendix:skewmodel} we consider a generalized version of Eq.~\ref{eqn:PV} that may be able to capture these features). Perhaps more important, self-gravity \citep[e.g.][]{federrath:2012.sfe.pwrspec.vs.time.sims}, large scale non-random perturbations in the ISM (spiral arms, mergers, etc.), and correlated feedback structures (radiation and SNe-induced ``bubbles'') can introduce very large deviations from Gaussianity \citep[although in simulations, it seems that a roughly normal PDF may still apply to each such ISM phase separately; see e.g.][]{hopkins:fb.ism.prop}. In any case, we are not arguing that the corrections here are necessarily most important in the ISM, but they may be critical to compare models and simulations, and to understand the behavior of even ideal turbulence.

\vspace{-0.7cm}
\acknowledgments 
We thank our referee, Christoph Federrath, for insightful discussions and critiques of this paper. Support for PFH was provided by NASA through Einstein Postdoctoral Fellowship Award Number PF1-120083 issued by the Chandra X-ray Observatory Center, which is operated by the Smithsonian Astrophysical Observatory for and on behalf of the NASA under contract NAS8-03060.\\

\bibliography{/Users/phopkins/Documents/work/papers/ms}

\clearpage

\begin{footnotesize}
\ctable[
  caption={{\normalsize Simulations Used in This Paper}\label{tbl:sims}},center,star
  ]{lcccccccccccc}{
\tnote[ ]{{\bf (1)} Reference: Source of simulation density PDF we fit (all data are taken directly from the published points in the PDFs). \\
{\bf (2)} Method: Numerical method. UG is uniform (Cartesian) grid; AMR is adaptive mesh refinement; SPH is smoothed-particle hydrodynamics. \\
{\bf (3)} Forcing: Turbulent velocity forcing scheme: C is purely compressive, S is solenoidal, M is mixed-mode driving \citep[see][]{federrath:2010.obs.vs.sim.turb.compare}. \\
{\bf (4)} Resolution: Numerical resolution. Refers to cell number for UG/AMR methods, particle number for SPH. \\
{\bf (5)} $\mathcal{M}$: Mach number (at the driving scale),  $\mathcal{M}=\langle v_{t}^{2} \rangle^{1/2}/c_{s}$. \\
{\bf (6)} $\mathcal{M}_{c}$: Typical (rms) compressive Mach number, where quoted by the authors. Otherwise, we assume this $=\mathcal{M}$ for compressive forcing, $=\mathcal{M}/3$ for solenoidal, and $=0.4\,\mathcal{M}$ for mixed-mode driving \citep[see][]{federrath:2010.obs.vs.sim.turb.compare,konstantin:mach.compressive.relation}. With magnetic fields, the ``effective'' compressive component is smaller than this.\\
{\bf (7)} $\mathcal{M}_{\rm A}$: Alf{\'e}n Mach number $\mathcal{M}_{\rm A}=\langle v_{t}^{2} \rangle^{1/2}/v_{\rm A}$. This is related to the ratio of thermal to magnetic pressure $\beta\equiv P_{\rm th}/P_{\rm mag} = 2\,c_{s}^{2}/v_{\rm A}^{2}$ by $\beta = 2\,(\mathcal{M}_{\rm A}/\mathcal{M})^{2}$. \\ 
{\bf (8)} $S_{\ln{\rho},\,V}$: Volume-weighted variance in $\ln{(\rho)}$ (computed directly from the simulation density distribution). \\ 
{\bf (9)} $S_{\ln{\rho},\,M}$: Mass-weighted variance in $\ln{(\rho)}$ (directly from the simulation density distribution). This is equal to $S_{\ln{\rho},\,V}$ in the lognormal case, but smaller if $T>0$. \\ 
{\bf (10)} $T$: Best-fit parameter $T$ from Eq.~\ref{eqn:PV}, describing the deviations from a normal PDF. The $1\,\sigma$ uncertainty is quoted, assuming all points in the PDF have equal logarithmic error bars (this is an upper limit to the uncertainty fitted to any of the data sets shown). \\
{\bf (11)} $T_{v}$: Best-fit parameter $T_{v}$ from Eq.~\ref{eqn:structfn.castaing} describing the intermittency in the velocity structure functions, where available (note that the structure functions for the \citealt{federrath:2010.obs.vs.sim.turb.compare} simulations are actually in \citealt{schmidt:2008.turb.structure.fns}). \\ 
{\bf (12)} $\delta_{\rm rms}^{T}$: Typical (rms) logarithmic deviation of the simulation data about the best-fit curve for the quoted $T$ value (in dex). \\
{\bf (13)} $\delta_{\rm rms}^{LN}$: Deviations as $\delta_{\rm rms}^{T}$, but for the best-fit lognormal model (Eq.~\ref{eqn:PV.LN}).
}
}{
\hline\hline
\multicolumn{1}{l}{Reference} &
\multicolumn{1}{c}{Method} & 
\multicolumn{1}{c}{Forcing} &
\multicolumn{1}{c}{Resolution} &
\multicolumn{1}{c}{$\mathcal{M}$} &  
\multicolumn{1}{c}{$\mathcal{M}_{c}$} &  
\multicolumn{1}{c}{$\mathcal{M}_{\rm A}$} &  
\multicolumn{1}{c}{$S_{\ln{\rho},\,V}$} &  
\multicolumn{1}{c}{$S_{\ln{\rho},\,M}$} &  
\multicolumn{1}{c}{$T$} &  
\multicolumn{1}{c}{$T_{v}$} &  
\multicolumn{1}{c}{$\delta_{\rm rms}^{T}$} &  
\multicolumn{1}{c}{$\delta_{\rm rms}^{LN}$} \\
\hline
\citet{federrath:2010.obs.vs.sim.turb.compare} & UG & C & $1024^{3}$ & $5.6$ & $5.5$ & $\infty$ & $9.23$ & $3.23$ & $0.42\pm0.06$ & $0.54\pm0.10$ & $0.045$ & $0.77$ \\ 
\citet{federrath:2010.obs.vs.sim.turb.compare} & UG & S & $1024^{3}$ & $5.3$ & $1.8$ & $\infty$ & $1.71$ & $1.45$ & $0.056\pm0.02$ & $0.17\pm0.03$ & $0.064$ & $0.42$ \\ 
\citet{schmidt:2009.isothermal.turb} & AMR & C & $768^{3}$ & $2.9$ & $2.5$ & $\infty$ & $3.33$ & $1.79$ & $0.23\pm0.03$ & $0.46\pm0.09$ & $0.055$ & $0.58$ \\ 
\citet{price:2010.grid.sph.compare.turbulence} & SPH & S & $512^{3}$ & $10$ & $3.3$ & $\infty$ & $2.24$ & $2.01$ & $0.036\pm0.014$ & $0.20\pm0.03$ & $0.14$ & $0.51$ \\ 
\citet{price:2010.grid.sph.compare.turbulence} & UG & S & $512^{3}$ & $10$ & $3.3$ & $\infty$ & $2.97$ & $1.90$ & $0.16\pm0.02$ & $0.27\pm0.07$ & $0.17$ & $1.23$ \\ 
\citet{kritsuk:2007.isothermal.turb.stats} & UG/AMR & M & $1024^{3}$ & $6$ & $2.2$ & $\infty$ & $1.30$ & $1.28$ & $0.003\pm0.017$ & $0.12\pm0.01$ & $0.078$ & $0.11$ \\ 
\hline
\citet{konstantin:mach.compressive.relation} & UG & C & $512^{3}$ & $15$ & $13.6$ & $\infty$ & $12.93$ & $4.65$ & $0.38\pm0.16$ & -- & $0.057$ & $0.58$ \\ 
\citet{konstantin:mach.compressive.relation} & UG & S & $512^{3}$ & $15$ & $7.4$ & $\infty$ & $4.58$ & $2.14$ & $0.25\pm0.08$ & -- & $0.11$ & $1.24$ \\ 
\citet{konstantin:mach.compressive.relation} & UG & C & $512^{3}$ & $5.5$ & $4.8$ & $\infty$ & $8.17$ & $3.10$ & $0.34\pm0.16$ & $0.28\pm0.05$ & $0.091$ & $0.89$ \\ 
\citet{konstantin:mach.compressive.relation} & UG & S & $512^{3}$ & $5.5$ & $2.4$ & $\infty$ & $1.77$ & $1.47$ & $0.064\pm0.033$ & $0.16\pm0.04$ & $0.039$ & $0.60$ \\ 
\citet{konstantin:mach.compressive.relation} & UG & C & $512^{3}$ & $2.0$ & $2.0$ & $\infty$ & $5.51$ & $2.26$ & $0.29\pm0.12$ & -- & $0.049$ & $0.95$ \\ 
\citet{konstantin:mach.compressive.relation} & UG & S & $512^{3}$ & $2.0$ & $0.95$ & $\infty$ & $0.68$ & $0.62$ & $0.023\pm0.024$ & -- & $0.078$ & $0.36$ \\ 
\citet{konstantin:mach.compressive.relation} & UG & C & $512^{3}$ & $0.5$ & $0.4$ & $\infty$ & $0.35$ & $0.27$ & $0.054\pm0.014$ & -- & $0.067$ & $0.74$ \\ 
\citet{konstantin:mach.compressive.relation} & UG & S & $512^{3}$ & $0.5$ & $0.05$ & $\infty$ & $0.019$ & $0.007$ & $0.018\pm0.003$ & -- & $0.21$ & $2.41$ \\ 
\citet{konstantin:mach.compressive.relation} & UG & C & $512^{3}$ & $0.1$ & $0.1$ & $\infty$ & $0.026$ & $0.023$ & $0.0001\pm0.0002$ & -- & $0.12$ & $0.12$ \\ 
\citet{konstantin:mach.compressive.relation} & UG & S & $512^{3}$ & $0.1$ & $0.0013$ & $\infty$ & $7.4$e-$5$ & $6.9$e-$6$ & $0.002\pm0.001$ & -- & $0.43$ & $5.86$ \\ 
\hline
\citet{kowal:2007.log.density.turb.spectra} & UG & S & $256^{3}$ & $7.1$ & $2.4$ & $7.1$ & $1.26$ & $1.23$ & $0.007\pm0.045$ & $0.12\pm0.02$ & $0.050$ & $0.094$ \\ 
\citet{kowal:2007.log.density.turb.spectra} & UG & S & $256^{3}$ & $2.34$ & $0.78$ & $7.4$ & $0.62$ & $0.55$ & $0.037\pm0.037$ & $0.10\pm0.03$ & $0.042$ & $0.21$ \\ 
\citet{kowal:2007.log.density.turb.spectra} & UG & S & $256^{3}$ & $0.74$ & $0.25$ & $7.4$ & $0.064$ & $0.059$ & $0.023\pm0.009$ & $0.14\pm0.04$ & $0.14$ & $0.42$ \\ 
\citet{kowal:2007.log.density.turb.spectra} & UG & S & $256^{3}$ & $0.36$ & $0.12$ & $7.2$ & $0.0053$ & $0.0051$ & $0.015\pm0.003$ & $0.15\pm0.04$ & $0.24$ & $1.23$ \\ 
\citet{kowal:2007.log.density.turb.spectra} & UG & S & $256^{3}$ & $7.0$ & $2.0$ & $0.70$ & $1.39$ & $1.30$ & $0.012\pm0.045$ & $0.11\pm0.01$ & $0.081$ & $0.19$ \\ 
\citet{kowal:2007.log.density.turb.spectra} & UG & S & $256^{3}$ & $2.20$ & $0.63$ & $0.69$ & $0.66$ & $0.60$ & $0.034\pm0.037$ & $0.076\pm0.006$ & $0.10$ & $0.22$ \\ 
\citet{kowal:2007.log.density.turb.spectra} & UG & S & $256^{3}$ & $0.68$ & $0.19$ & $0.65$ & $0.076$ & $0.068$ & $0.038\pm0.014$ & $0.074\pm0.006$ & $0.23$ & $0.66$ \\ 
\citet{kowal:2007.log.density.turb.spectra} & UG & S & $256^{3}$ & $0.33$ & $0.09$ & $0.64$ & $0.010$ & $0.0097$ & $0.024\pm0.005$ & $0.057\pm0.008$ & $0.29$ & $0.93$ \\ 
\hline
\citet{molina:2012.mhd.mach.dispersion.relation} & UG & M & $256^{3}$ & $17.6$ & $7.04$ & $\infty$ & $3.68$ & $2.11$ & $0.32\pm0.07$ & -- & $0.078$ & $0.57$ \\ 
\citet{molina:2012.mhd.mach.dispersion.relation} & UG & M & $256^{3}$ & $16.8$ & $6.72$ & $7.0$ & $2.59$ & $1.90$ & $0.049\pm0.024$ & -- & $0.13$ & $0.23$ \\ 
\citet{molina:2012.mhd.mach.dispersion.relation} & UG & M & $256^{3}$ & $10.6$ & $4.24$ & $\infty$ & $2.89$ & $1.84$ & $0.27\pm0.08$ & -- & $0.10$ & $0.58$ \\ 
\citet{molina:2012.mhd.mach.dispersion.relation} & UG & M & $256^{3}$ & $10.2$ & $4.08$ & $9.0$ & $2.16$ & $1.62$ & $0.049\pm0.030$ & -- & $0.17$ & $0.29$ \\ 
\citet{molina:2012.mhd.mach.dispersion.relation} & UG & M & $256^{3}$ & $5.4$ & $2.16$ & $\infty$ & $1.81$ & $1.37$ & $0.12\pm0.04$ & -- & $0.075$ & $0.43$ \\ 
\citet{molina:2012.mhd.mach.dispersion.relation} & UG & M & $256^{3}$ & $4.98$ & $1.99$ & $8.4$ & $1.35$ & $1.06$ & $0.076\pm0.026$ & -- & $0.14$ & $0.43$ \\ 
\citet{molina:2012.mhd.mach.dispersion.relation} & UG & M & $256^{3}$ & $2.21$ & $0.88$ & $\infty$ & $0.59$ & $0.50$ & $0.058\pm0.022$ & -- & $0.12$ & $0.48$ \\ 
\citet{molina:2012.mhd.mach.dispersion.relation} & UG & M & $256^{3}$ & $2.09$ & $0.84$ & $8.1$ & $0.52$ & $0.43$ & $0.076\pm0.018$ & -- & $0.17$ & $0.74$ \\ 
\hline\hline\\
}
\end{footnotesize}

\clearpage

\begin{appendix}

\section{A Derivation from a Simple Random Counts-in-Cells Model}
\label{sec:appendix:deriv}

Consider the following simple, phenomenological model for stochastic variation in densities as a function of scale. Begin with a box containing total mass $M_{0}$ and volume $V_{0}$, hence mean (volume-average) density $\rho_{0}=M_{0}/V_{0}$; without loss of generality choose units so $M_{0}=V_{0}=\rho_{0}=1$. We then take a series of ``steps.'' In each step, divide the volume $V$ into two equal parts $=V/2$ (it does not matter how we perform this division). Select one of the volumes, and repeat until we reach a final volume $V_{f}$ corresponding to the scale on which we wish to measure the densities. 

Consider two classes of steps. In a ``normal'' step, the density is smooth on the scale of interest, so each sub-volume contains exactly $1/2$ the mass initially $M$ within the volume $V$ and density is conserved. In a ``variable'' step, there is some inhomogeneous structure within the volume that makes the density non-uniform; consider the simple model where we assign to one sub-volume ($V/2$) a random fraction $f_{i}$ of the mass $M$ (where $f_{i}$ is a uniform random variable between $0\le f_{i}\le 1$). By mass conservation, the other volume has mass fraction $\tilde{f}_{i}=1-f_{i}$. But $1-f_{i}$ is {\em also} distributed as a uniform random variable between $0-1$. So for any final volume, selected at random, the mass is distributed as $2^{-m}\,\prod_{i=1}^{i=n}\,f_{i}$, where $m$ is the number of ``normal'' steps and $n$ the number of ``variable'' steps. The final volume is $=2^{-(m+n)}$, so the final density is distributed as $\rho=2^{n}\,\prod_{i=1}^{i=n}\,f_{i}$ (independent of $m$). So 
\begin{align}
\ln{(\rho)} &= n\,\ln{2} + \sum_{i=0}^{i=n}\,\ln{(f_{i})} \\ 
P{\Bigl(}u_{n}=\sum_{i=0}^{i=n}\,\ln{(f_{i})}{\Bigr)} &= P(\ln{f_{1}}) \otimes P(\ln{f_{2}}) \otimes ... \otimes P(\ln{f_{n}}) \\ 
&= \frac{1}{\Gamma(n)}\,|u_{n}|^{n-1}\,\exp{(-|u_{n}|)}
\end{align}
where $\otimes$ denotes convolution. The latter equality follows because, for a uniform random variate $f$ ($P(f)\,{\rm d}f={\rm d}f$, $0\le f\le 1$), it is trivial to show $P(\ln{f})\,{\rm d}\ln{f} = \exp{(-|\ln{f}|)}\,{\rm d}\ln{f}$ for $\ln{f}<0$. So we just need the convolution over $n$ independent exponentially distributed variables, i.e.\ a gamma distribution. 

If the distribution of inhomogeneous structures is non-uniform, then the number $n$ of ``variable'' steps is itself random; but assume they are statistically homogeneous on large scales, with a mean value $\langle n \rangle = \lambda$ (for whatever final scale we average on) then this is just a counting process and $n$ is Poisson-distributed. Combining, we obtain the final distribution 
\begin{align}
P(\ln{\rho}) = \sum_{n=0}^{\infty}\frac{\lambda^{n}\,e^{-\lambda}}{n!}\,
\frac{(-\ln{\rho} + n\ln{2})^{n-1}}{\Gamma(n)}\,e^{(\ln{\rho} - n\ln{2})}
\end{align}
where the sum is restricted to $\ln{\rho} < n\,\ln{2}$ for each $n$. 

But this is clearly very closely related to our proposed Eq.~\ref{eqn:PV}, for the choice $T=1$. In fact, \citet{castaing:1996.thermodynamic.turb.cascade} showed that this is simply the ``quantized'' version of the PDF in Eq.~\ref{eqn:PV} (i.e.\ has $x_{0}=\ln{2}>0$ in the notation therein). This is because we have considered discrete steps in scale, with quantized ``random'' fluctuations (i.e.\ a binary choice between a multiplication by $f_{i}$ or no change). If we instead allowed continuous steps in ``scale'' (i.e.\ differential steps in the volume element) but enforce the same ``rate,'' per logarithmic change in volume, of encountering such a multiplicative ``event'' that divides the mass into $f_{i}$ and $1-f_{i}$, then we simply recover exactly the derivation in \citet{castaing:1996.thermodynamic.turb.cascade} for $T=1$ and, as a result,
\begin{align}
P(\ln{\rho}) = \sum_{n=0}^{\infty}\frac{\lambda^{n}\,e^{-\lambda}}{n!}\,
\frac{(-\ln{\rho} + \lambda/2)^{n-1}}{\Gamma(n)}\,e^{(\ln{\rho} - \lambda/2)}
\end{align}
for $\ln{\rho}\le\lambda/2$ ($P(\ln{\rho})=0$ otherwise). We can think of this as essentially taking the $n$ ``outside'' of the sum (replacing it with $\lambda$), since the number of ``steps'' in log-volume is always fixed (independent of the number of ``random'' steps) going from the box scale to the measurement scale.

This is not a unique model for the density distribution, of course, but it {\em by construction} obeys mass conservation and exactly describes the density PDF on all scales, i.e.\ resolving the problems in \S~\ref{sec:intro}. We also stress that this is not the only model which arrives at the $T=1$ density PDF; however it provides a useful context in which to interpret the PDF, analogous to the use of the discrete $\beta$ model to motivate and interpret the intermittency model of \citet{sheleveque:structure.functions}. 

We can further generalize this model to different $T$ (at least for $T\ll1$). In a ``variable'' step, consider dividing the volume into $T^{-1}+1$ equal sub-volumes and taking all but one of them (a fraction $T^{-1}/(T^{-1}+1) = 1/(1+T)$) as the ``total sub-volume'' in the step. Assume each sub-volume has a random fraction of the mass, subject to mass conservation as before, such that the mass-fraction PDF for any randomly-selected one of the sub-volumes is identical (again as before). For $T\ll1$, any such distribution leads to a total mass fraction in the $T^{-1}$ chosen sub-volumes distributed as $f^{T}$ to leading order (where $f$ is again a uniform variate between $0-1$; this must be true because the PDF of the sum over $1/T$ subvolumes -- product of their individual probabilities -- must converge to an $f$-like distribution). The final volume is then $2^{-m}\,(1+T)^{-n}$, and the final mass is distributed as $2^{-m}\,\Pi\,f_{i}^{T}$, so the logarithm of this is distributed as $2^{-m}\,\sum\ln{(f_{i}^{T})} = 2^{-m}\,T\,\sum{\ln{(f_{i})}}$. Thus we obtain 
\begin{align}
P(\ln{\rho}) &= T^{-1}\,\sum_{n=0}^{\infty}\frac{\lambda^{n}\,e^{-\lambda}}{n!}\,
\frac{x^{n-1}}{\Gamma(n)}\,e^{-x} \\ 
x &= T^{-1}[-\ln{\rho} + n\,\ln{(1+T)}]
\end{align}
which is again just the quantized version of our proposed density PDF (following \citealt{castaing:1996.thermodynamic.turb.cascade} to make this continuous with a constant ``rate'' parameter leads to exactly our proposed Eq.~\ref{eqn:PV}).

In this interpretation, then, $\lambda$ is equal to the mean number of ``events'' or ``structures'' producing density inhomogeneities -- point processes in the cascade -- integrated over a range of scales from the driving scale down to the averaging scale. To the extent that the density distribution converges, it is because these processes vanish below some scale (becoming less significant below the sonic scale, and completely vanishing by the viscous scale). $T$ represents the typical fractional variation associated with one such event (or equivalently the fraction of the mass on a given scale ``associated with'' a typical event). Because the density fluctuations associated with individual typical shocks, rarefactions, etc., are fractionally larger as the compressive Mach number increases, we naturally expect a growing $T$ with $\mathcal{M}_{c}$.

\vspace{-0.5cm}
\section{A More General Model Allowing Different Skewness}
\label{sec:appendix:skewmodel}

The cascade described above was derived by assuming a one-sided scaling in $\ln{\rho}$. But as noted in \citet{castaing:1996.thermodynamic.turb.cascade}, mathematically speaking the same derivation could trivially generalize to fluctuations of opposite sign in $\ln{\rho}$ (though this may not be physical). Consider each fluctuation to have either positive or negative sign (in log space) with probability of positive fluctuations $p_{+}$. 
This gives a PDF:
\begin{align}
 P(u)\,{\rm d}u &= \sum_{n=0}^{\infty}\,P_{\lambda}(n)\,
 \sum_{m=0}^{n}\, P_{+}(m\,|\,n) \,
P_{T}(u\,|\,n,\,m)\,{\rm d}u \\ 
\nonumber P_{\lambda}(n) &= \frac{\lambda^{n}\,e^{-\lambda}}{n!} \\ 
\nonumber P_{+}(m\,|\,n) &= \frac{n!}{m!\,(n-m)!}\,p_{+}^{m}\,(1-p_{+})^{n-m} \\ 
\nonumber P_{T}(u\,|\,n,\,m) &= \frac{|u|^{n-1}\,e^{-|u|}}{(m-1)!\,(n-m-1)!}\,H(u\,|\,n,\,m) \\ 
\nonumber H(u\,|\,n,\,m) &= 
   \left\{ \begin{array}{ll}
      \int_{0}^{\infty}{\rm d}t\,e^{-2\,|u|\,t}\,t^{n-m-1}\,(1+t)^{m-1} & (u\ge0) \\ 
      \int_{0}^{\infty}{\rm d}t\,e^{-2\,|u|\,t}\,t^{m-1}\,(1+t)^{n-m-1} & (u<0) \
\end{array}
    \right. \\ 
\nonumber u_{0}(\lambda,\,T) &= -\frac{1}{T}\,\ln{{\Bigl[}
\int_{-\infty}^{+\infty}{\rm d}u\,\exp{(T\,u)}\,P(u)
{\Bigr]}} \\ 
\nonumber \rho &= \exp{[T\,(u+u_{0})]}\,\rho_{0}
\end{align}
Here, $P_{\lambda}$ is the Poisson probability of $n$ events; $P_{+}$ is the binomial probability of $m$ ``positive-biased'' events (with per-event probability $p_{+}$); $P_{T}$ is the convolution of the exponential distribution of amplitudes for a continuous-time relaxation process with $m$ positive-amplitude events and $n-m$ negative amplitude events; and $u_{0}$ is determined by mass conservation:
\be
\frac{1}{\rho_{0}}\int{\rm d}u\,{\rho}\,P(u)=\int_{-\infty}^{+\infty}\,{\rm d}u\,\exp{[T\,(u+u_{0})]}\,P(u) = 1
\ee

The difference between this and the formulation in the text is that here, the choice of $p_{+}$ determines the level of skewness in the log-density PDF ($p_{+}=0$ corresponds to our model in the text). If we choose $p_{+}=1$ we simply ``mirror'' the skewness in $\ln{\rho}$. The choice $p_{+}=1/2$ is non-skew -- although still non-Gaussian at a desired level set by $T$ -- since positive and negative perturbations are symmetric. This may be useful for fitting non-isothermal cases: for example, for a polytropic gas with $c_{s}^{2}\propto \rho^{\gamma-1}$ and $\gamma<1$, the PDF can be skewed towards higher densities \citep[see][]{scalo:1998.turb.density.pdf,passot:1998.density.pdf}, represented here by $p_{+}>1/2$.

\end{appendix}

\end{document}